\documentclass[journal,twoside,web]{ieeecolor}
\usepackage{etoolbox}
\makeatletter
\@ifundefined{color@begingroup}%
{\let\color@begingroup\relax
\let\color@endgroup\relax}{}%
\def\fix@ieeecolor@hbox#1{%
\hbox{\color@begingroup#1\color@endgroup}}
\patchcmd{\@makecaption}{\hbox}{\fix@ieeecolor@hbox}{}{\FAILED}
\patchcmd{\@makecaption}{\hbox}{\fix@ieeecolor@hbox}{}{\FAILED}

\usepackage{makecell}

\usepackage{generic}
\usepackage{amsmath,amssymb,amsfonts}
\usepackage{graphicx}
\definecolor{myblue}{RGB}{51,158,225}

\usepackage[colorlinks=true, linkcolor=blue, citecolor=blue]{hyperref}
\usepackage[capitalize]{cleveref}

\def\BibTeX{{\rm B\kern-.05em{\sc i\kern-.025em b}\kern-.08em
    T\kern-.1667em\lower.7ex\hbox{E}\kern-.125emX}}
\begin{document}
\title{Development of 3D Pixel Sensors via an 8-inch CMOS-Compatible Process}
\author{Huimin Ji, Zhihua Li, Wenzheng Cheng, Zheng Li, Kai Huang, Jing Wen, Song Liu, Manwen Liu, and Jun Luo
\thanks{This work was supported in part by the National Key Research and Development Program of China under Grant 2023YFF0719600 and in part by the
National Natural Science Foundation of China under Grant 12375188. (Corresponding author: Manwen Liu.) }
\thanks{Huimin Ji, Zhihua Li, Song Liu, Manwen Liu, and Jun Luo are with the Key Laboratory of Fabrication Technologies for Integrated Circuits, Chinese Academy of Sciences, Beijing 100029, also with the Institute of Microelectronics, Chinese Academy of Sciences, Beijing 100029, China, and also with the School of Integrated Circuits, University of Chinese Academy of Sciences, Beijing 100049, China (e-mail: liumanwen@ime.ac.cn).}
\thanks{Wenzheng Cheng, Kai Huang, and Jing Wen are with the Key Laboratory of Fabrication Technologies for Integrated Circuits, Chinese Academy of Sciences, Beijing 100029, and also with the Institute of Microelectronics, Chinese Academy of Sciences, Beijing 100029, China.}
\thanks{Zheng Li is with the School of Integrated Circuits, Ludong University, Yantai 264025, China, and also with the Institute of Microelectronics, Chinese Academy of Sciences, Beijing 100029, China.}
}

\maketitle

\begin{abstract}
In the construction of High-Luminosity Large Hadron Collider (HL-LHC) and Future Circular Collider (FCC) experiments, 3D pixel sensors have become indispensable components due to their superior radiation hardness, fast response, and low power consumption. However, there are still significant challenges in the process of 3D sensors manufacturing. In this work, single devices and arrays of 3D sensors based on 30 $\mu$m epitaxial silicon wafer have been designed, simulated, fabricated, and tested. This process was developed on the 8-inch CMOS process platform of the Institute of Microelectronics of the Chinese Academy of Sciences (IMECAS). The key processes include Deep Reactive Ion Etching (DRIE) with the Bosch process, in-situ doping, and an innovative back-etching. After testing the 3D pixel sensors, we have summarized the leakage current and capacitance of devices with different sizes with respect to bias voltages. We also found that the fabricated devices were almost all successfully produced, which laid a strong foundation for subsequent large-scale mass production.
\end{abstract} 


\section{Introduction}
\label{sec:introduction}
\IEEEPARstart{T}{he} relentless pursuit of high-performance radiation detectors for particle physics experiments, medical imaging, and space applications has driven the evolution of semiconductor sensor technologies. Among these, the 3D pixel sensors proposed by S. Parker \emph{et al.} in 1997 \cite{b1} represent a paradigm shift from traditional planar designs. In this design, electrodes penetrate the silicon substrate perpendicular to the surface. This approach offers superior radiation hardness \cite{b2}, fast charge collection, and low power consumption. These advantages become particularly crucial for future collider experiments facing extreme radiation environments.

The RD50 collaboration pioneered systematic radiation tolerance studies of 3D sensors by establishing standardized testing protocols under high-fluence conditions \cite{b5}, laying the foundation for their first large-scale application in the ATLAS Insertable B-Layer (IBL) project, marking the technology's successful deployment at the Large Hadron Collider (LHC) \cite{b6}. Research teams from the Barcelona Institute of Microelectronics (IMB-CNM) \cite{b7, b8, b9}, Fondazione Bruno Kessler (FBK) \cite{b10, b11, b12}, and SINTEF \cite{b13, b14, b15, b16} have also made significant breakthroughs in process innovation.

Despite these advancements, significant fabrication difficulties persist in 3D pixel sensors on production scale, including electrode shape control, aspect ratio, doping and filling processes, wafer flatness, mechanical stress, and warpage. The complex 3D electrode structures require advanced semiconductor processes, including deep reactive ion etching (DRIE) with the Bosch process \cite{b3} and in-situ doping. The intricate interplay between electrode geometry and electrical performance requires TCAD simulation-guided optimization \cite{b4}. Moreover, the long manufacturing cycle of 6 to 12 months and high costs further increase the development difficulty.

Fabrication of 3D sensors utilizing CMOS process is significantly important to large scale manufacture and integration to ASIC and readout circuit. To fulfill this, we designed and fabricated 3D pixel sensors on the 8-inch CMOS process platform in the Institute of Microelectronics of the Chinese Academy of Sciences (IMECAS). Fabricated 3D pixel sensors include single devices and array devices. We successfully developed a trench/column electrode structure with an aspect ratio of 70:1 (0.5 $\mu$m width), significantly increasing the sensor fill factor. This 3D trench/column structure effectively enhances the electrical insulation performance between adjacent pixels, while also featuring low depletion voltage and fast signal acquisition capability. The proposed 3D trench/column sensors are based on mature electrode theoretical frameworks \cite{b17, b18}, that provide a reliable theoretical foundation for design. In this work, we will disclose more details of the fabrication process of 3D pixel sensors, providing a guarantee for large-scale production in the future.

The organizational structure of the remaining part of this paper is as follows: Section II elaborates in detail on the simulation process and results of single device of the 3D pixel sensors using the TCAD tool. Section III describes the designed fabrication process and key process innovations with process details. Section IV presents leakage current and capacitance tests and analyses of the completed wafers, and then Section V presents a preliminary summary of this work with an outlook of future research directions.

\section{Device Modeling and Simulation}
\begin{figure}[!t]
\centerline{\includegraphics[width=\columnwidth]{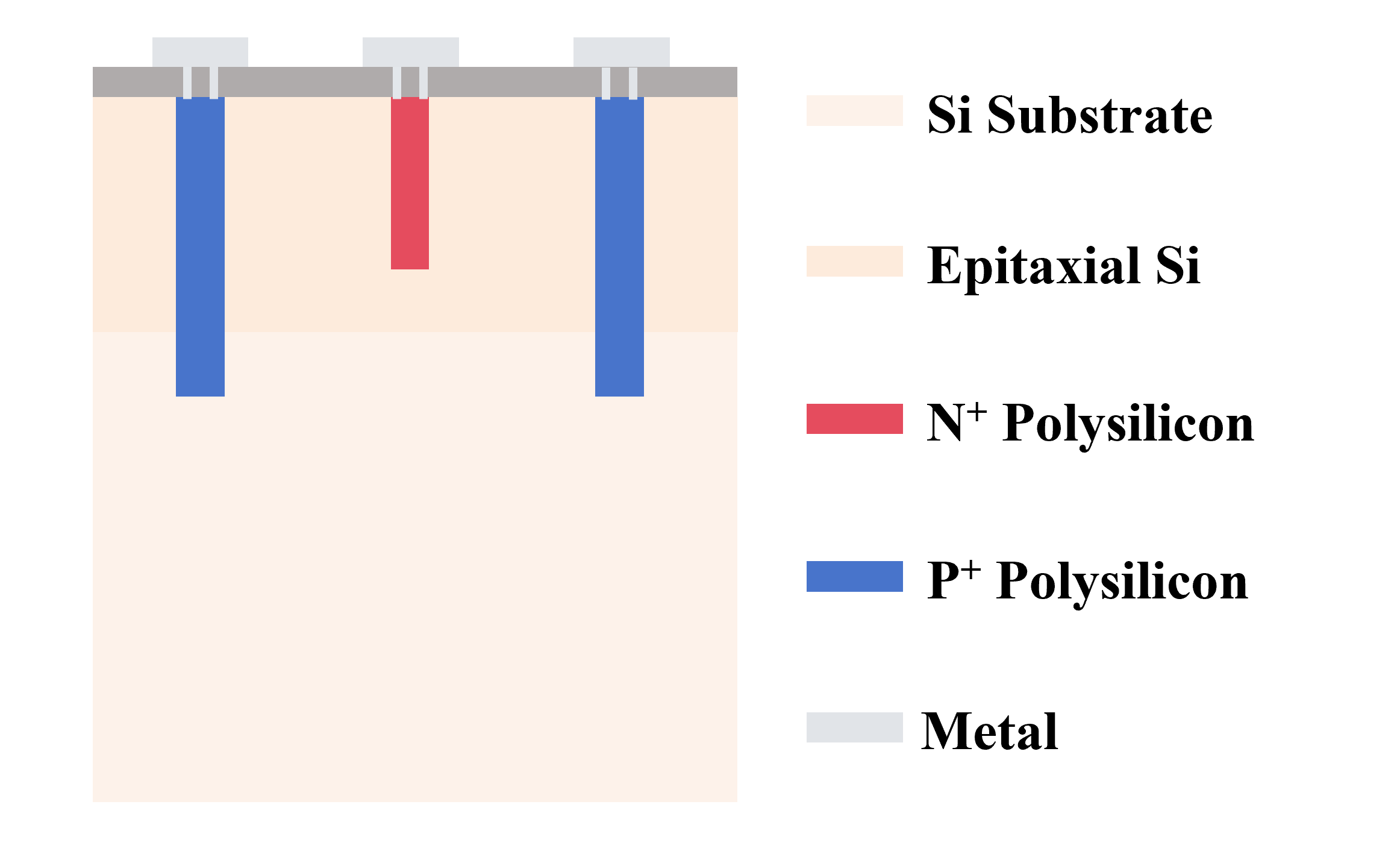}}
\caption{Schematic of the single device structure.}
\label{fig: 2D schematic}
\end{figure}

\begin{figure}[!t]
\centerline{\includegraphics[width=\columnwidth]{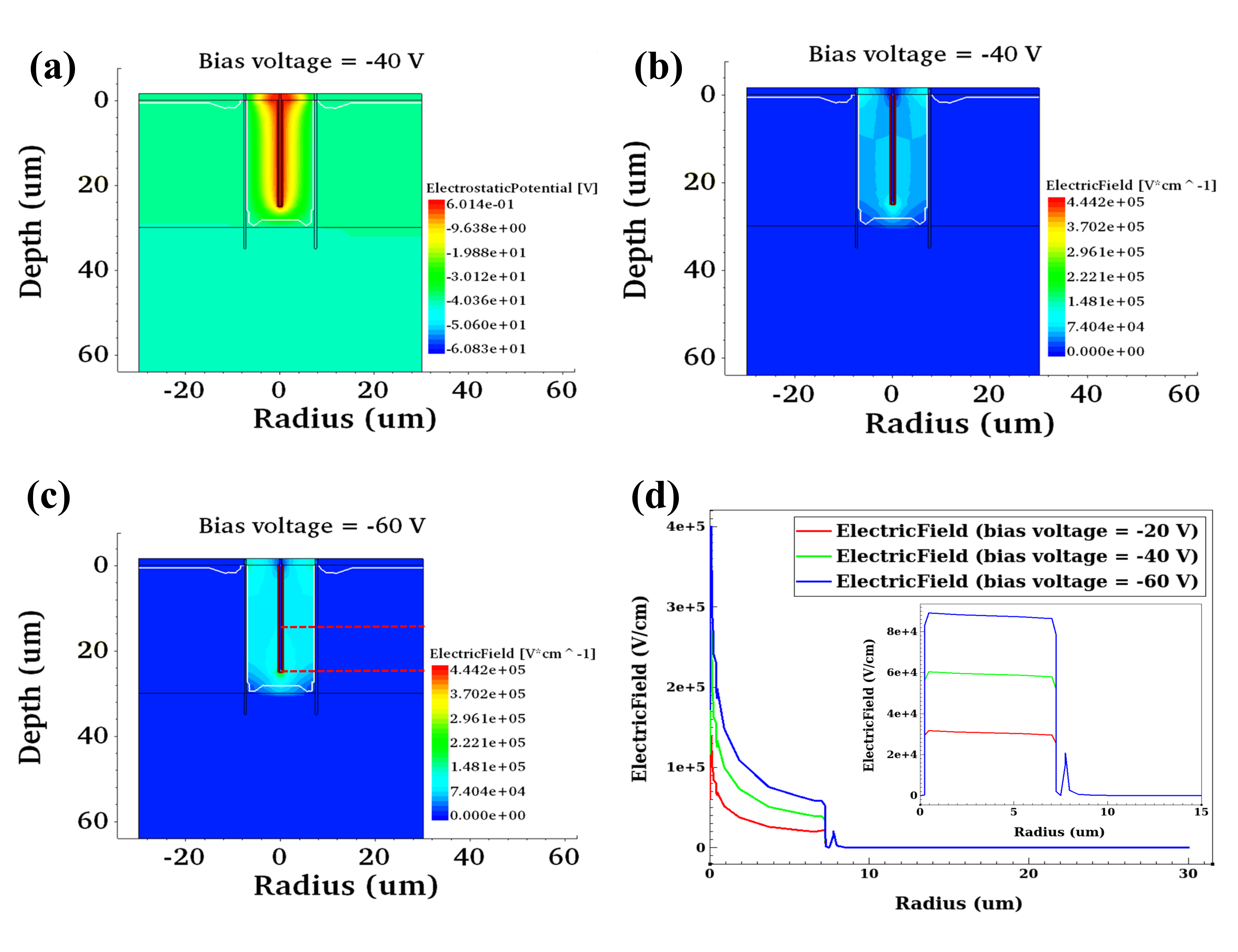}}
\caption{
      \textcolor{myblue}{(a)} The potential distribution with a reverse bias voltage of 40 V. 
      \textcolor{myblue}{(b)} The electric field distribution with a reverse bias voltage of 40 V. 
      \textcolor{myblue}{(c)} The electric field distribution with a reverse bias voltage of 60 V.
      \textcolor{myblue}{(d)} The one-dimensional transverse electric field distribution at the depths of 25 $\mu$m and 15 $\mu$m for the device under different bias voltages.}
\label{fig: electric}
\end{figure}

\cref{fig: 2D schematic} shows the 2D structure schematic of a single 3D sensor device. The device was fabricated on a 30 $\mu$m high-resistance epitaxial silicon, with the central column electrode doped with phosphorus at about 4$\times  10^{20}$ atoms/cm$^{-3}$. The side trench electrode that surrounding the central column electrode is doped with boron at about 2.8$\times  10^{21}$ atoms/cm$^{-3}$. The central column electrode is functioned as the collection electrode. The trench electrode is used to apply reverse bias voltage.

We used the TCAD tool to construct this structure and simulated the electric field and potential distributions of the device under different bias voltages. \cref{fig: electric}\textcolor{blue}{(a)} and \textcolor{blue}{(b)} show the electric field and potential distribution when the bias voltage is -40 V. \cref{fig: electric}\textcolor{blue}{(c)} demonstrates the electric field when the bias voltage is -60 V. We found that as the bias voltage increases, the electric field in the depletion region of the device becomes stronger. \cref{fig: electric}\textcolor{blue}{(d)} presents the electric field distribution at the bottom of the collection electrode, with an inset illustrating the electric field profile in the middle of the collection electrode. The exact location is shown by the dotted lines in \cref{fig: electric}\textcolor{blue}{(c)}. The analysis reveals a significant concentration of high electric field intensity adjacent to the bottom of the central collection electrode. It is attributed to the fact that the PN junction is located near the central electrode. By comparison, it can be seen that the electric field at the bottom of the central collection electrode is the highest. Under a 60 V bias, the electric field at the bottom of the collection electrode reaches 4$\times  10^{5}$ V/cm. It is sufficient to induce impact ionization and may lead to device breakdown, which is consistent with the subsequent test results.

\begin{figure}[!t]
\centerline{\includegraphics[width=\columnwidth]{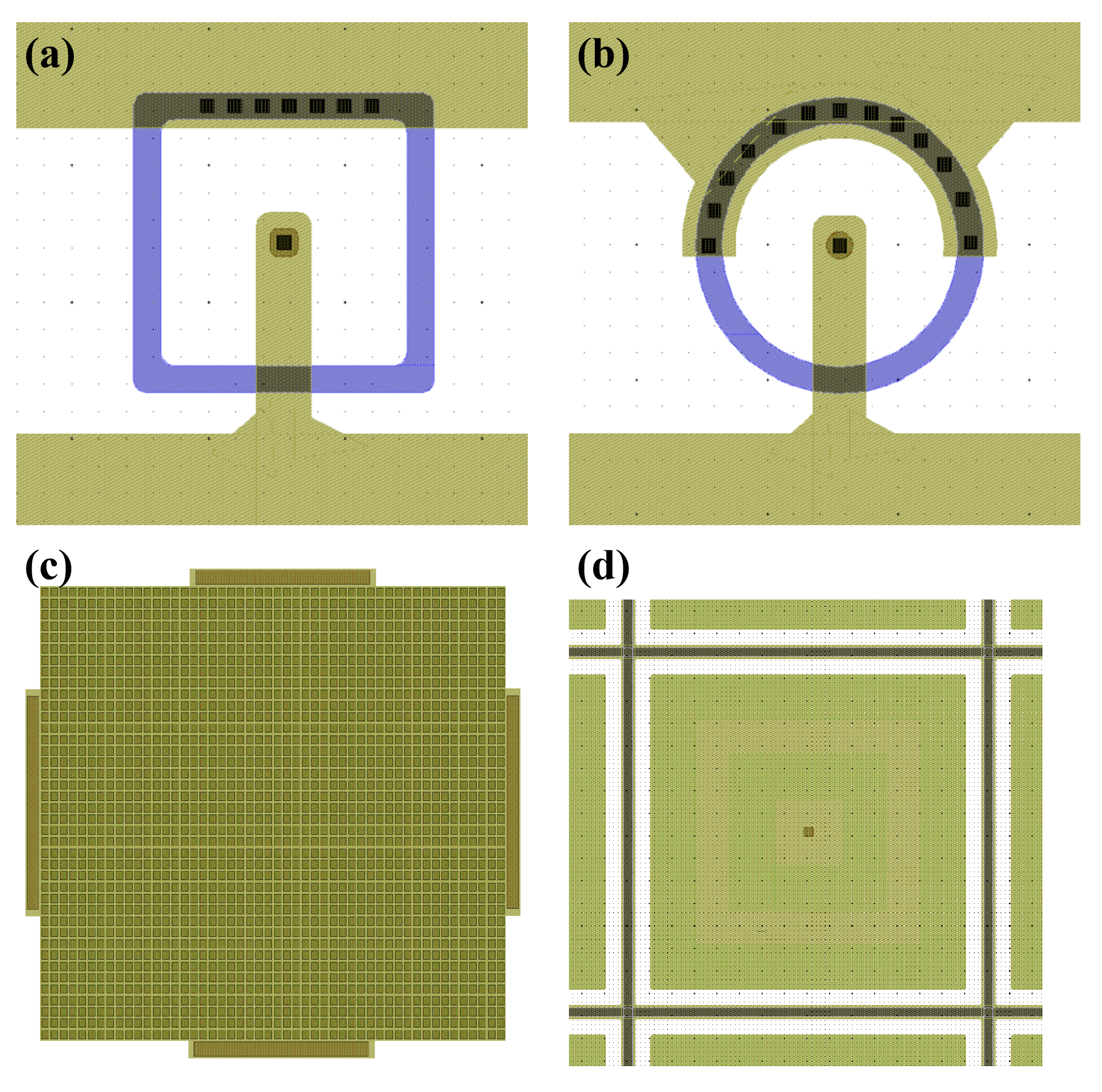}}
 \caption{
        Layouts of the 
      \textcolor{myblue}{(a)} square single device, 
      \textcolor{myblue}{(b)} circular single device, 
      \textcolor{myblue}{(c)} array with a pixel size of 80 $\mu$m $\times$ 80 $\mu$m, and
      \textcolor{myblue}{(d)} single device in the array with a pixel size of 80 $\mu$m $\times$ 80 $\mu$m
        of 3D pixel sensors.
    }
\label{fig: layout}
\end{figure}

\cref{fig: layout} shows the layouts of two individual devices (square and circular) and the array with a pixel size of 80 $\mu$m $\times$ 80 $\mu$m of 3D pixel sensors. All our subsequent chip fabrication processes were carried out based on these layouts.

\section{Fabrication Process}
\begin{figure}[!t]
\centerline{\includegraphics[width=\columnwidth]{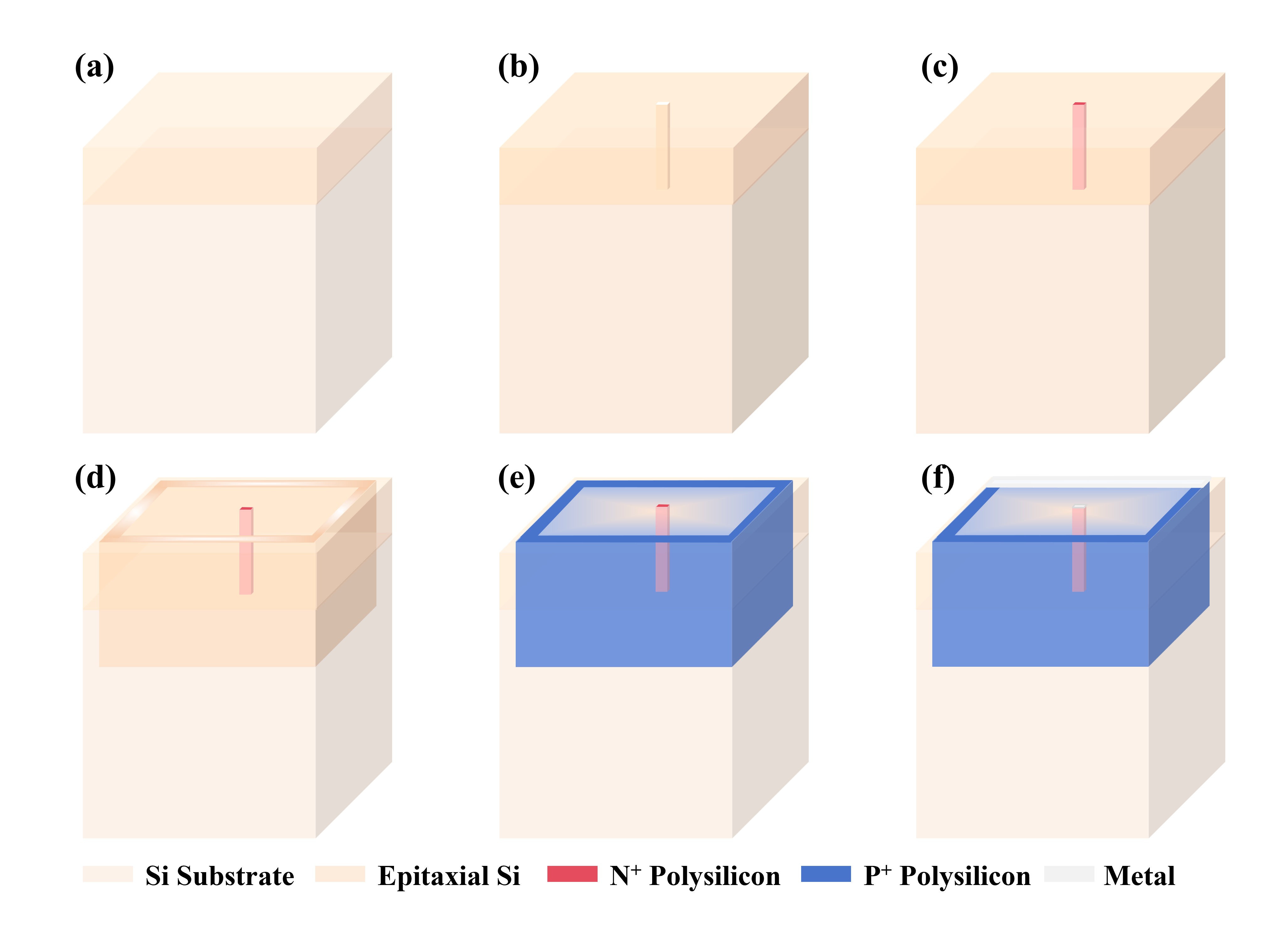}}
\caption{Fabrication processes of the single device of 3D pixel sensors.}
\label{fig: fabprocess}
\end{figure}

We will take the single device shown in \cref{fig: fabprocess} as an example to introduce the basic fabrication processes of the 3D pixel sensors. First, we prepare a high-resistance epitaxial silicon wafer with an epitaxial thickness of 30 $\mu$m as shown in \cref{fig: fabprocess}\textcolor{blue}{(a)}. Next, at the center of each device, we use DRIE with the Bosch process to create column with a width of 0.5--2 $\mu$m and a depth of 20--25 $\mu$m as shown in \cref{fig: fabprocess}\textcolor{blue}{(b)}. We use the corresponding thickness of photoresist based on the feature size of the pattern. After the dielectric etching, the remaining photoresist and the oxide layer are utilized as a hard mask for deep silicon etching. Then, we used in-situ doping to dope it into an N-type as shown in \cref{fig: fabprocess}\textcolor{blue}{(c)}. Similarly, at the periphery of each device, we used DRIE with the Bosch process to create surrounding trench with a width of 0.5--2 $\mu$m and a depth of 30--50 $\mu$m as shown in \cref{fig: fabprocess}\textcolor{blue}{(d)}.  Then, we use in-situ doping to dope it into a P-type as shown in \cref{fig: fabprocess}\textcolor{blue}{(e)}. Finally, the  fabrication is completed by making metal electrodes for contact as shown in \cref{fig: fabprocess}\textcolor{blue}{(f)}.

\begin{figure}[!t]
\centerline{\includegraphics[width=\columnwidth]{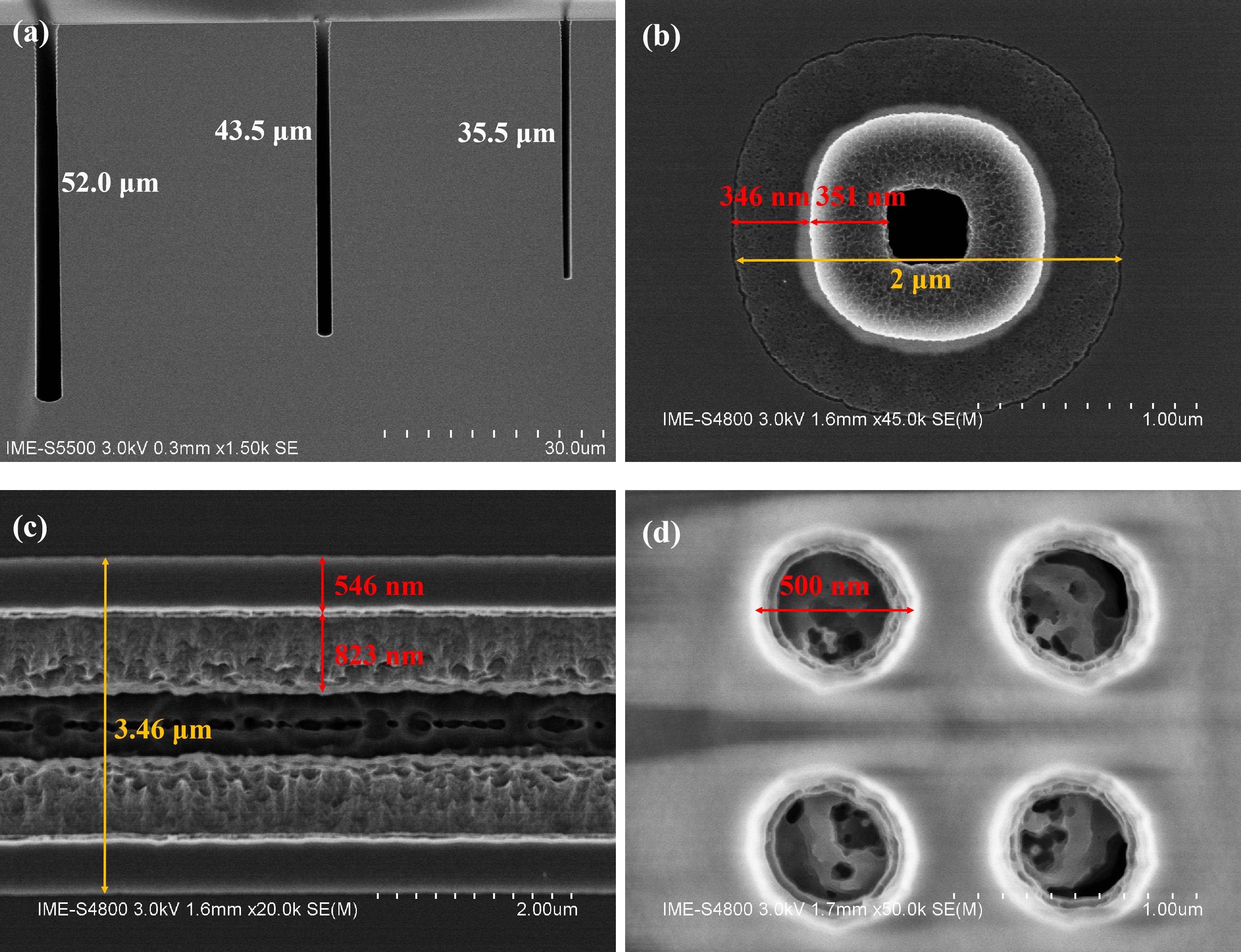}}
\caption{The images of the trenches of 3D pixel sensors by using ultra high resolution scanning electron microscope (S-5500).
      \textcolor{myblue}{(a)} Etching depth corresponding to different trench/column width under the same etching time. 
      \textcolor{myblue}{(b)} Surface image of the back-etching process of polysilicon on the 3D electrode. 
      \textcolor{myblue}{(c)} The cross-sectional view of the 3D trench after the back-etching process of polysilicon on the 3D electrode.
      \textcolor{myblue}{(d)} The surface image of the central electrode vias etching.
        }
\label{fig: SEM of trench}
\end{figure}

Next, we will introduce the specific details of the fabrication processes. The 3D trench/column electrodes have high verticality after etching. This is because during the DRIE with the Bosch process, we always use an oxide layer or photoresist as a hard mask. The thickness of the hard mask should be set reasonably according to the number of etching cycles and the overall process flow. Moreover, for different trench/column widths, the etching depth is different under the same etching time. Under the same etching time, the wider the trench/column width, the deeper the etching depth, as shown in \cref{fig: SEM of trench}\textcolor{blue}{(a)} \cite{b19}. We obtain an aspect ratio of 70:1 when the trench/column width is 0.5 $\mu$m.

After 3D trench/column etching, doped polysilicon is deposited via Chemical Vapor Deposition (CVD) using phosphorus or boron as the dopant source. This is a doping technique carried out during the material growth process, which can avoid damaging the crystal structure. We usually refer to this as in-situ doping. To prevent short circuits, the surface polysilicon of the 3D electrode needs to be completely removed after deposition. We adopt the back-etching process with a high selectivity to stop at the oxide layer, ensuring that the electrode surface is clean and free of residue, while the polysilicon on the sidewalls of the 3D trench/column electrodes is retained. The SEM image of the surface after the back-etching process is shown in \cref{fig: SEM of trench}\textcolor{blue}{(b)}, and the SEM image of the side of the 3D trench/column electrode after the back-etching process is shown in \cref{fig: SEM of trench}\textcolor{blue}{(c)}. As can be seen from the picture, the surface is smooth and flat, which lays the foundation for the subsequent contact fabrication.

After completing the polysilicon back-etching process, we deposit and planarize an oxide layer to fill voids inside the polysilicon and seal surface openings. This step enhances the insulation between electrodes and prevents surface leakage, which is critical for improving the stability and reliability of the 3D pixel sensors. Subsequently, we use photolithography and development to transfer the via pattern onto the wafer and ensure that the developed vias align with the underlying polysilicon. Then, we used dry etching technology to remove the oxide layer areas that were not protected by the photoresist, thereby forming vertical vias. We also checked the results to confirm the exposure status of the polycrystalline silicon on the silicon wafer. As shown in \cref{fig: SEM of trench}\textcolor{blue}{(d)}, it is the SEM image after the via etching.

Before the metal deposition, we use hydrofluoric acid (HF) to remove the residual oxide on the surface of polysilicon and then deposit Ti/TiN. This can reduce the contact resistance and increase the adhesion between tungsten and the oxide layer. Then, we perform isotropic deposition of the tungsten film to completely fill the vias. Similar to the back-etching process of polysilicon, we use back-etching technology to remove excess tungsten after vias filling to prevent short circuits between P$^{+}$ and N$^{+}$ electrodes. After completing tungsten etching, we perform over-etching to further remove the underlying TiN and Ti layers, thereby finishing the back-etching process.

\begin{figure}[!t]
\centerline{\includegraphics[width=\columnwidth]{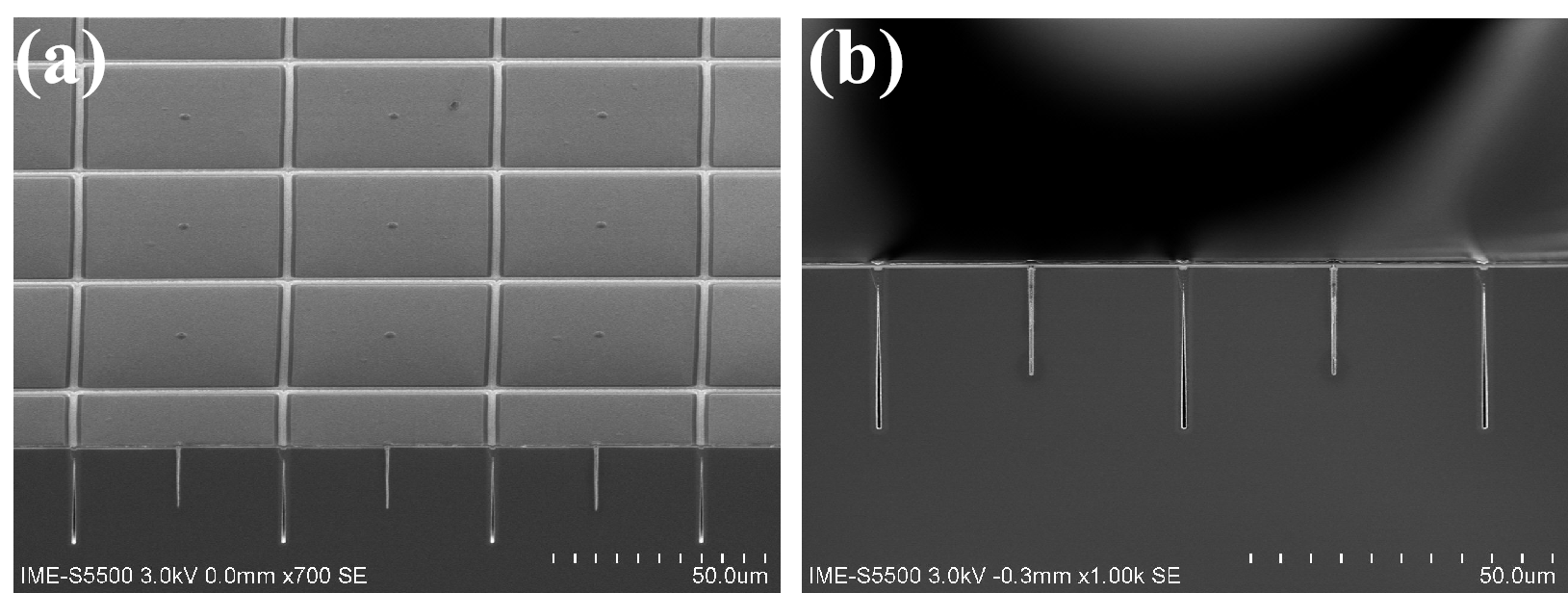}}
\caption{\textcolor{myblue}{(a)}The overall view and \textcolor{myblue}{(b)}side view of the filled trenches and columns for the completed array.}
\label{fig: S5500 of trench}  
\end{figure}

\cref{fig: S5500 of trench} shows the surface and side profiles of the trenches and columns in the 3D pixel sensors that have been completed. It can be seen that all the trenches and columns have been filled and the surface is smooth.

\begin{figure}[!t]
\centerline{\includegraphics[width=\columnwidth]{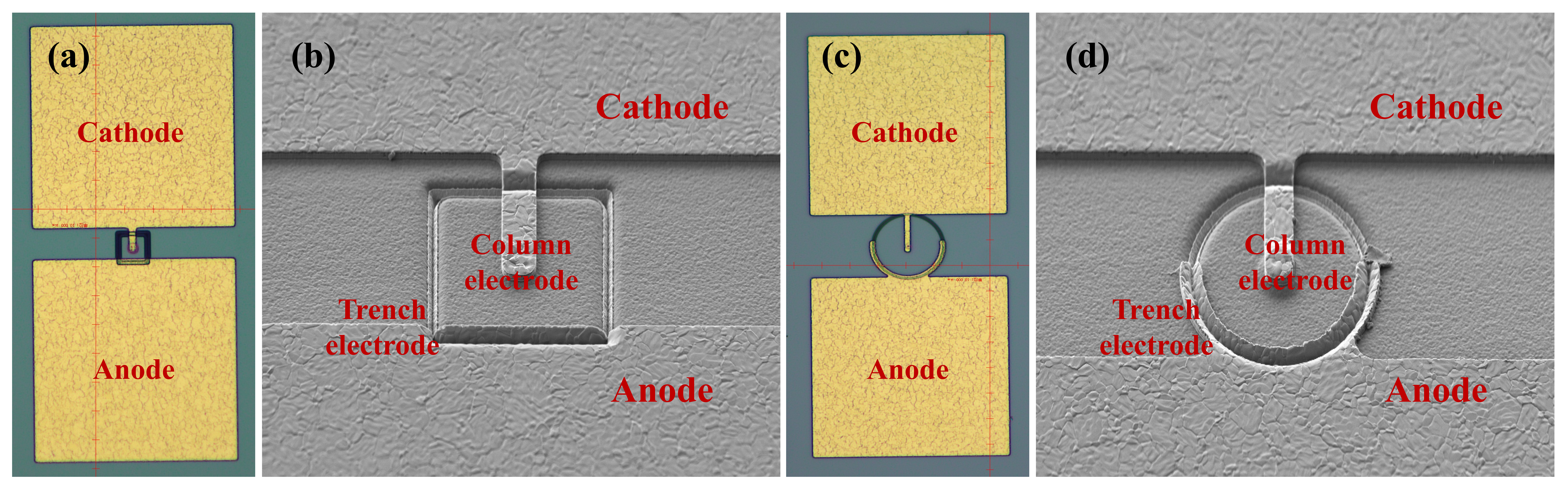}}
\caption{\textcolor{myblue}{(a)} Microscope view and 
\textcolor{myblue}{(b)} Focused Ion Beam Microscope (FIB) view of the square shape of the single device of 3D pixel sensor.
\textcolor{myblue}{(c)} Microscope view and 
\textcolor{myblue}{(d)} Focused Ion Beam Microscope (FIB) view of the circular shape of the single device of 3D pixel sensor.
}
\label{fig: electrode shape}
\end{figure}

As shown in \cref{fig: electrode shape}, the physical images of two types of IMECAS fabricated 3D pixel sensors with trench shapes are presented under microscope and Focused Ion Beam Microscope (FIB). The electrode is composed of polysilicon and metal. The anode marked in the figure is the power supply electrode, where a negative reverse bias voltage is applied. The cathode marked in the figure is the read-out electrode where a 0 V voltage will be applied during the test.

\begin{figure}[!t]
\centerline{\includegraphics[width=\columnwidth]{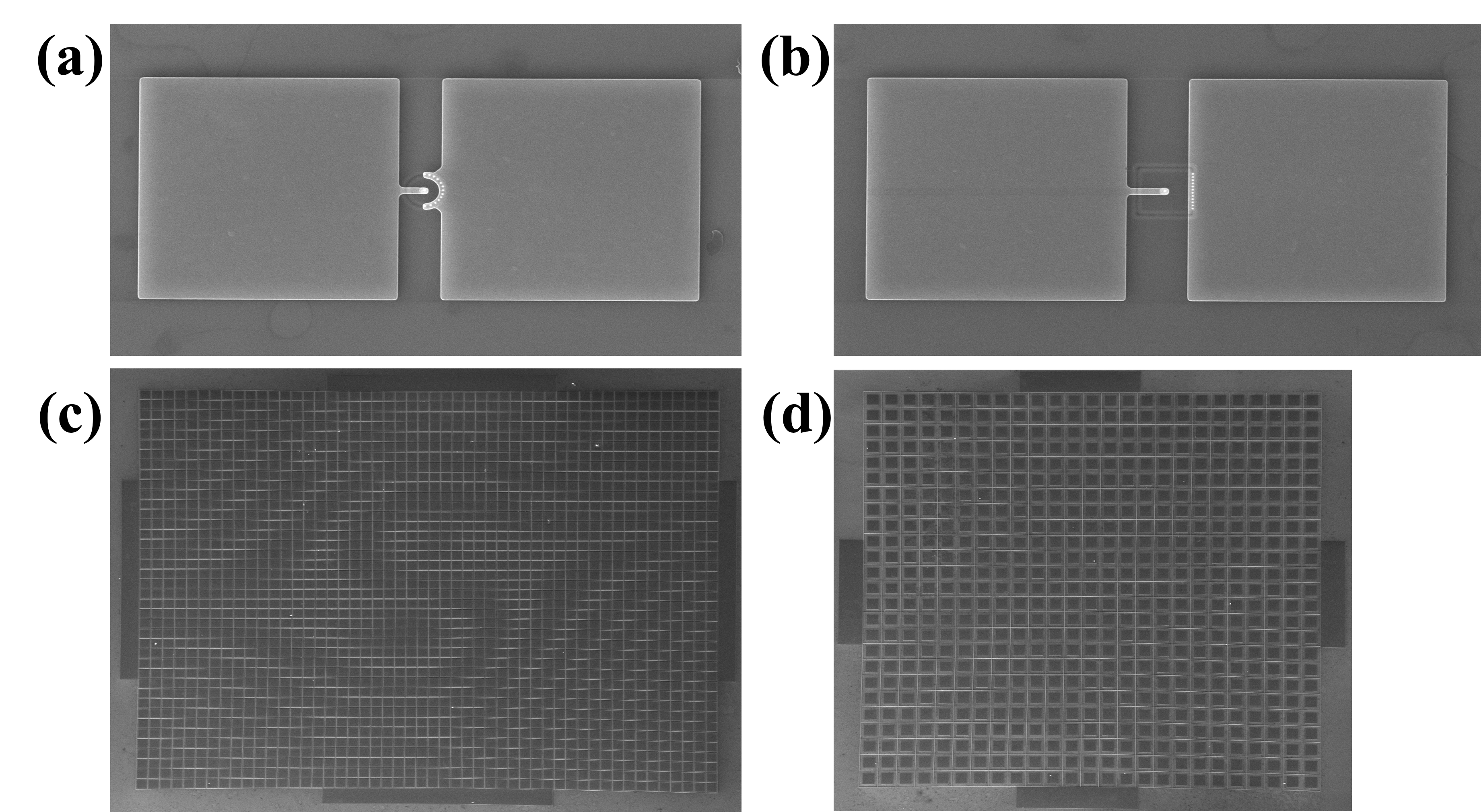}}
\caption{The surface morphology of 
      \textcolor{myblue}{(a)} the single device with circular shape,
      \textcolor{myblue}{(b)} the single device with square shape,
      \textcolor{myblue}{(c)} the array with pixel size of 150 $\mu$m $\times$ 150 $\mu$m, and 
      \textcolor{myblue}{(d)} the array with pixel size of 80 $\mu$m $\times$ 80 $\mu$m 
      of 3D pixel sensors observed with Scanning Electron Microscope (SEM).
      }
\label{fig: TESCAN}
\end{figure}

\cref{fig: TESCAN} presents the surface morphology of the single devices and arrays of 3D pixel sensors observed with Scanning Electron Microscope (SEM). The single devices of 3D pixel sensors in circular and square shapes are shown in \cref{fig: TESCAN}\textcolor{blue}{(a)} and \cref{fig: TESCAN}\textcolor{blue}{(b)} respectively. The array with pixel size of 150 $\mu$m $\times$ 150 $\mu$m is shown in \cref{fig: TESCAN}\textcolor{blue}{(c)}. The array with pixel size of 80 $\mu$m $\times$ 80 $\mu$m is shown in \cref{fig: TESCAN}\textcolor{blue}{(d)}. Here, we merely selected devices with different shapes for display. In fact, we have also pulled offsets of different sizes for devices of the same shape. We will summarize these parameters as well as the test results in the following tables (\cref{array_pixel_parameters} and \cref{single_pixel_parameters}).

\section{Measurements and Analysis}
\begin{figure}[!t]
\centerline{\includegraphics[width=\columnwidth]{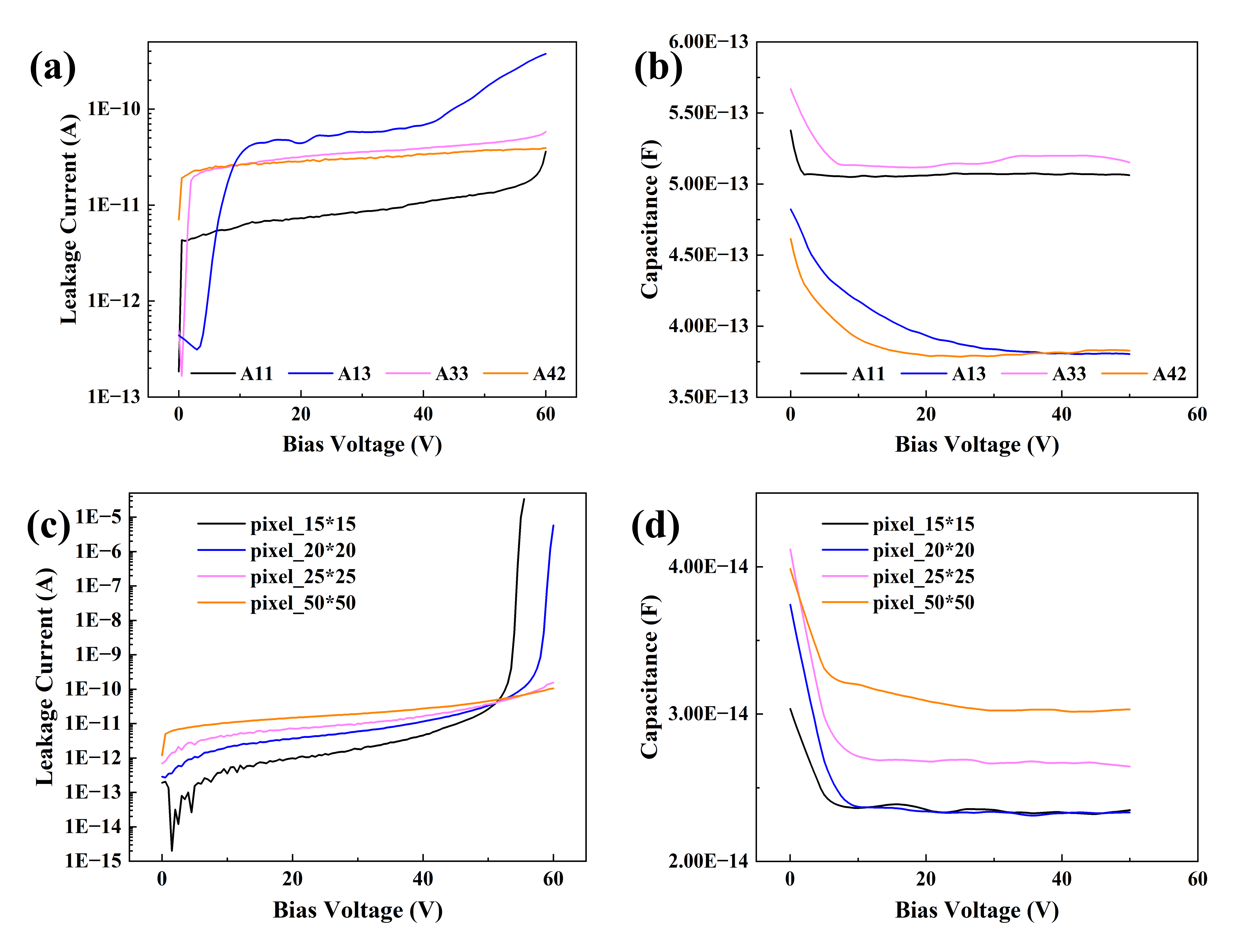}}
\caption{The test results of the arrays and single devices of 3D pixel sensors.
      \textcolor{myblue}{(a)} The curves of leakage current of four arrays of 3D pixel sensors as a function of voltage. 
      \textcolor{myblue}{(b)} The curves of capacitance of four arrays of 3D pixel sensors as a function of voltage.
      \textcolor{myblue}{(c)} The curves of leakage current of four single devices of 3D pixel sensors with different pixel sizes varying with voltage.
      \textcolor{myblue}{(d)} The curves of the capacitance of four single devices of 3D pixel sensors with different pixel sizes varying with voltage.
      }
\label{fig: test}
\end{figure}

\begin{table*}
\centering
\caption{Parameters of the Arrays of 3D Pixel Sensors}
\label{array_pixel_parameters}
\setlength{\tabcolsep}{3pt} 
\begin{tabular}{llllll}
\hline
\textbf{Name} & 
\textbf{Pixel Size (\boldmath$\mu$m$^{2}$)} & 
\textbf{Array Dimensions} & 
\textbf{Trench/Column Width (\boldmath$\mu$m)} & 
\textbf{Leakage Current (A) @ 20 V} & 
\textbf{Capacitance (F) @ 20 V} \\
\hline
A11 & 80 $\times$ 80 & 40 $\times$ 50 & 2.0 & 7.32 $\times  10^{-12}$ & 5.06 $\times  10^{-13}$  \\
A13 & 150 $\times$ 150 & 40 $\times$ 50 & 1.0 & 4.29 $\times  10^{-11}$ & 3.94 $\times  10^{-13}$ \\
A33 & 80 $\times$ 80 & 40 $\times$ 50 & 1.0 & 3.18 $\times  10^{-11}$ & 5.12 $\times  10^{-13}$  \\
A42 & 150 $\times$ 150 & 40 $\times$ 50 & 2.0 & 2.84 $\times  10^{-11}$ & 3.79 $\times  10^{-13}$ \\
\hline
\end{tabular}
\end{table*}

\begin{table*}
\centering
\caption{Parameters of the single devices of 3D Pixel Sensors}
\label{single_pixel_parameters}
\setlength{\tabcolsep}{3pt} 
\begin{tabular}{llllll}
\hline
\textbf{Name} & 
\textbf{Pixel Size (\boldmath$\mu$m$^{2}$)} & 
\textbf{Shape} & 
\textbf{Trench/Column Width (\boldmath$\mu$m)} & 
\textbf{Leakage Current (A) @ 20 V} & 
\textbf{Capacitance (F)} \\
\hline
DL(1--5) & 10 $\times$ 10 & \makecell[l]{Square (DL1, 2, 3) \\Circular (DL4, 5)} & \makecell[l]{0.5 (DL1) \\ 1.0 (DL2, 4)\\ 2.0 (DL3, 5))} & 
\makecell[l]{DL1: 5.73 $\times  10^{-13}$  \\ DL2: 1.15 $\times  10^{-12}$ \\ DL3: 9.21 $\times  10^{-8}$ \\ DL4: 1.62 $\times  10^{-12}$ \\ DL5: 1.64 $\times  10^{-9}$} & 
\makecell[l]{DL1: 2.59 $\times  10^{-14}$ @ 20 V  \\ DL2: --- \\ DL3: 2.60 $\times  10^{-14}$ @ 5 V \\ DL4: 2.27 $\times  10^{-14}$ @ 5 V \\ DL5: 2.10 $\times  10^{-14}$ @ 10 V} \\

DL(6--8) & 35 $\times$ 35 & Square & \makecell[l]{0.5 (DL6) \\ 1.0 (DL7)\\ 2.0 (DL8))} & 
\makecell[l]{DL6: 1.04 $\times  10^{-11}$  \\ DL7: --- \\ DL8: 2.26 $\times  10^{-9}$} & 
\makecell[l]{DL6: 2.52 $\times  10^{-14}$ @ 20 V  \\  DL7: 6.78 $\times  10^{-15}$ @ 20 V \\ DL8: 1.25 $\times  10^{-14}$ @ 20 V} \\

DL(9--11) & 50 $\times$ 50 & Square & \makecell[l]{0.5 (DL9) \\ 1.0 (DL10)\\ 2.0 (DL11))} & 
\makecell[l]{DL9: 1.48 $\times  10^{-11}$  \\ DL10: 3.59 $\times  10^{-13}$ \\ DL11: 1.71$ \times  10^{-11}$} & 
\makecell[l]{DL9: 3.06 $\times  10^{-14}$ @ 20 V  \\  DL10: 3.08 $\times  10^{-14}$ @ 20 V \\ DL11: 4.70 $\times  10^{-14}$ @ 20 V} \\

DL(12--13) & 80 $\times$ 80 & Square & \makecell[l]{1.0 (DL12)\\ 2.0 (DL13))} & 
\makecell[l]{DL12: ---  \\ DL13: 1.01 $\times  10^{-11}$} & 
\makecell[l]{DL12: 5.05 $\times  10^{-14}$ @ 10 V  \\  DL13: 5.49 $\times  10^{-14}$ @ 10 V} \\

DL(14--16) & 150 $\times$ 150 & Square & \makecell[l]{0.5 (DL16) \\ 1.0 (DL15)\\ 2.0 (DL14))} & \makecell[l]{DL14: 4.87 $\times  10^{-11}$  \\ DL15: --- \\ DL16: 1.64 $\times  10^{-11}$} & 
\makecell[l]{DL14: 1.82 $\times  10^{-12}$ @ 10 V  \\  DL15: 1.29 $\times  10^{-12}$ @ 10 V \\ DL16: 1.26 $\times  10^{-12}$ @ 10 V} \\

DR(1--5) & 15 $\times$ 15 & \makecell[l]{Square (DR1, 2, 3) \\Circular (DR4, 5)} & 
\makecell[l]{0.5 (DR1) \\ 1.0 (DR2, 5)\\ 2.0 (DR3, 4))} & 
\makecell[l]{DR1: 9.79 $\times  10^{-13}$  \\ DR2: 1.65 $\times  10^{-12}$ \\ DR3: 2.50 $\times  10^{-9}$ \\ DR4: 5.02 $\times  10^{-10}$ \\ DR5: ---} & 
\makecell[l]{DR1: 2.38 $\times  10^{-14}$ @ 20 V  \\ DR2: 2.82 $\times  10^{-14}$ @ 20 V \\ DR3: --- \\ DR4: 3.38 $\times  10^{-14}$ @ 10 V \\ DR5: 2.85 $\times  10^{-15}$ @ 10 V} \\

DR(6--10) & 20 $\times$ 20 & \makecell[l]{Square (DR6, 7, 8) \\Circular (DR9, 10)} & 
\makecell[l]{0.5 (DR6) \\ 1.0 (DR7, 9)\\ 2.0 (DR8, 10))} & 
\makecell[l]{DR6: 3.70 $\times  10^{-12}$  \\ DR7: 1.80 $\times  10^{-12}$ \\ DR8: 6.05 $\times  10^{-8}$ \\ DR9: --- \\ DR10: ---} & 
\makecell[l]{DR6: 2.33 $\times  10^{-14}$ @ 20 V  \\ DR7: --- \\ DR8: 1.97 $\times  10^{-14}$ @ 10 V \\ DR9: 1.25 $\times  10^{-14}$ @ 10 V \\ DR10: 1.63 $\times  10^{-14}$ @ 10 V} \\

DR(11--15) & 25 $\times$ 25 & \makecell[l]{Square (DR11, 12, 13) \\Circular (DR14, 15)} & 
\makecell[l]{0.5 (DR11) \\ 1.0 (DR12, 14)\\ 2.0 (DR13, 15))} & 
\makecell[l]{DR11: 7.14 $\times  10^{-12}$  \\ DR12: --- \\ DR13: 9.24 $\times  10^{-8}$ \\ DR14: 3.64$ \times  10^{-12}$ \\ DR15: 1.01 $\times  10^{-8}$} & 
\makecell[l]{DR11: 2.72 $\times  10^{-14}$ @ 20 V  \\ DR12: 1.78 $\times  10^{-14}$ @ 10 V \\ DR13: --- \\ DR14: 2.97 $\times  10^{-12}$ @ 10 V \\ DR15: 2.21 $\times  10^{-14}$ @ 10 V} \\

\hline
\end{tabular}
\end{table*}

As summarized in \cref{array_pixel_parameters}, we can see geometry parameters and I-V/C-V results of four different arrays of 3D pixel sensors. The I-V and C-V curves are shown in \cref{fig: test}\textcolor{blue}{(a)} and \cref{fig: test}\textcolor{blue}{(b)}. Through analysis of the measurement results, it is observed that sensor arrays featuring smaller pixel dimensions generally exhibit reduced leakage current. The capacitance is dependent on the width and the depth of the collecting electrode theoretically. The measured capacitance values range from approximately 300 fF to 500 fF. The comparatively low leakage current and small capacitance are attributed to the use of a thin epitaxial layer as well as innovations in the fabrication process.

\cref{single_pixel_parameters} summarizes geometry parameters and I-V/C-V results of thirty-one different single devices of 3D pixel sensors. We selected four devices (DR1, DR6, DR11, and DL9) with the same shape and trench/column width but different pixel sizes to plot their I-V and C-V curves, as shown in \cref{fig: test}\textcolor{blue}{(c)} and \cref{fig: test}\textcolor{blue}{(d)}. We observed that as the pixel size increased, the leakage current and capacitance of the single device also increased. By comparing DL9, DL10, and DL11 in \cref{single_pixel_parameters}, we found that among those with the same shape and pixel size but trench/column widths of 0.5 $\mu$m, 1.0 $\mu$m, and 2.0 $\mu$m, the single device DL10 with a trench/column width of 1.0 $\mu$m had the lowest leakage current, whereas the capacitance increased with the trench/column width.

Furthermore, systematic analysis of the \cref{single_pixel_parameters} reveals a distinct correlation between pixel sizes and leakage current characteristics in devices with 2 $\mu$m trench/column width. For smaller pixel sizes, such as 10 $\mu$m $\times$ 10 $\mu$m, 15 $\mu$m $\times$ 15 $\mu$m, 20 $\mu$m $\times$ 20 $\mu$m, 25 $\mu$m $\times$ 25 $\mu$m, and 35 $\mu$m $\times$ 35 $\mu$m, the single-device leakage current demonstrates significantly higher values. In contrast, devices with larger pixel sizes, such as 50 $\mu$m $\times$ 50 $\mu$m, 80 $\mu$m $\times$ 80 $\mu$m, and 150 $\mu$m $\times$ 150 $\mu$m, exhibit a remarkable reduction in leakage current by approximately 1 to 2 orders of magnitude when incorporating the 2 $\mu$m trench/column width configuration. This may be caused by the fact that for small pixels, surface current may become dominant.

\section{Conclusion}
In this article, we have elaborated on the entire processes of designing, simulating, layouting, fabricating and testing of 3D pixel sensors. The performance evaluation results clearly demonstrate that the sensors fabricated at IMECAS exhibit outstanding characteristics, such as the leakage current of most 3D pixel sensors less than 10 pA at 20V, and the capacitance below 3 pF. These attributes are critical for applications in high-radiation environments such as the HL-LHC upgrades. It can be conclusively demonstrated that we have successfully realized functional 3D pixel sensors with high yield and performance uniformity. More importantly, the expertise acquired in key steps such as DRIE with the Bosch process of high-aspect-ratio holes, polysilicon electrode filling, and back-etching process has enabled proficient mastery of the core fabrication processes. This achievement marks an important milestone in the development of radiation sensor technology, and lays a solid foundation for the future large-scale production of 3D pixel sensors using the CMOS platform.

\end{document}